\documentclass[a4paper,12pt]{article}
\usepackage{times}
\usepackage{geometry}
\usepackage{amsthm,amsmath,amssymb}
\usepackage[utf8]{inputenc}

\newtheorem{proposition}{Proposition}

\newtheorem{theorem}{Theorem}

\newtheorem{remark}{Remark}

\usepackage{mathrsfs}
\usepackage{color}
\usepackage{graphicx}
\usepackage{subfigure}
\usepackage{float}
\usepackage{caption}
\usepackage{multirow}
\usepackage{graphicx}
\usepackage{epstopdf}
\usepackage{booktabs}
\usepackage{makecell}
\usepackage{overpic}

\usepackage[colorlinks, linkcolor = blue, anchorcolor = blue, urlcolor = blue, citecolor = blue]{hyperref}
\usepackage{authblk}

\newcommand\R{\mathbb{R}}
\newcommand{\Ro}{\mathcal{R}_0}
\newcommand\andrea[1]{#1}

\newcommand\keywords[1]{\textbf{Keywords}: #1}
\title{On the effectiveness of odor-baited traps\\ on mosquito-borne infections} 
\author[1]{Haonan Zhong}
\author[2]{Marina Lima}
\author[2]{Andrea Pugliese}
\author[2]{Michela Sabbatino}
\author[2]{Cinzia Soresina}
\author[2,3]{Tamas Tekeli}
\affil[1]{School of Mathematics and  Statistics, Southwest University, Chongqing, PR China}
\affil[2]{Department of Mathematics, University of Trento, Trento, Italy}
\affil[2,3]{Bolyai Institute, University of Szeged, Szeged,  Hungary}
\date{}                     
\begin{document}
\maketitle
\abstract{
    The host's odor serves as a critical biological tracking signal in the host-seeking process of mosquitoes, and its heterogeneity significantly influences the transmission of mosquito-borne diseases. In this study, we propose a mosquito-borne disease model incorporating odor-baited traps to examine the impact of odor on disease transmission. Following recent experimental evidence, we also assume that infected humans are more attractive than susceptible or recovered ones.
    The value of the basic reproduction number, $\Ro$, depends on the attractiveness of traps, adjusted relative to infected individuals; the dependence on the relative attractiveness of susceptibles is non-monotone, suggesting that there exists an optimal mosquito preference that maximizes disease transmission. When $\Ro>1$, there exists an endemic equilibrium which, under certain conditions, is proved to be globally stable. An endemic equilibrium may also exist when $\Ro < 1$, due to a backward bifurcation occurring when infected humans incur significant mortality. The phenomenon of backward bifurcation is reduced when odor-baited traps are more abundant.
    Analytical results and simulations show that deploying traps and enhancing their lethality for mosquitoes can help reduce disease prevalence and the risk of an outbreak. However, the attenuation of odor in highly attractive traps may lead to a rebound in the epidemic, especially when the traps gradually lose their attractiveness compared to susceptible hosts.

    }
    

\section{Introduction}	
Mosquito-borne infections are diseases transmitted through the bites of infected mosquitoes carrying pathogens such as bacteria, viruses, and parasites~\cite{WHOa}. Major mosquito-borne diseases include malaria, dengue, West Nile virus, yellow fever, Chikungunya, Japanese encephalitis, o Zika, all characterized by high morbidity and mortality rates~\cite{WHOa, Lee2018, Chala2021}.
Globalization, urbanization, and agricultural development have allowed these diseases to spread beyond tropical regions and reach global scales~\cite{Chala2021}. 
Malaria, caused by parasites and spread by \textit{Anopheles} mosquitoes, resulted in 2023 in an estimated 263 million cases and almost 600,000 deaths, primarily among children under five~\cite{WHOc}. 
Dengue, transmitted by \textit{Aedes} mosquitoes, is the most common mosquito-borne viral infection. Its worldwide incidence  has grown dramatically 
in recent decades, with the number of reported cases increasing from 505,430 cases in 2000 to 14.6 million in 2024~\cite{WHOb}. This number underestimates the number of new infections, since many are asymptomatic or self-managed. It has been estimated that in 2017 there were approximately 106 million cases, causing around 40,000 deaths~\cite{Zeng2021,WHOb,Lee2018,Franklinos2019,Chala2021,Zeng2021}.
According to the World Health Organization (WHO), vector-borne diseases account for more than 17\% of all infectious diseases globally, with mosquito-borne diseases being the largest contributors to the burden, and pose a significant global public health challenge~\cite{WHOa}.

Several of these infections, like dengue~\cite{dengue_europe} or chikungunya~\cite{chik_world, Wang2025}, have been spreading to new areas in subtropical or temperate climates, due to increased globalization and climate change~\cite{George2024}. Mathematical models have been developed to understand the threat posed worldwide by mosquito-borne infections~\cite{Felipe2021, Zardini2024}.
	
As well known, mosquito-borne diseases cannot spread directly between humans, but their transmission depends on vectors, specifically infected female mosquitoes. Mosquito-to-human transmission occurs when an infectious female mosquito feeds, after mating, on human blood to acquire the nutrients necessary for egg production, and so injects the pathogen into the host along with saliva during the bite, leading (with some probability) to infection~\cite{WHOa,Lee2018,Tolle2009}.
Human-to-mosquito transmission instead occurs when a susceptible female mosquito bites an infected host and may acquire the pathogens. 

This two-stage process is the basis of mathematical modelling of mosquito-borne infections~\cite{APSDR_rev}, starting from the pioneering work of Ronald Ross~\cite{Ross1910}. Beyond discovering the cycle of malaria transmission, his work has led to the basis of the concept of the reproduction ratio $\Ro$, from which it follows that control measures capable of bringing $\Ro$ below 1 are sufficient to eradicate malaria (or other mosquito-borne infections). His model was extended by Macdonald~\cite{Macdonald1957}, who acknowledged the relevance of incubation periods (delays) between the moment of infection and the start of infectiousness. What is known as the Ross--Macdonald model~\cite{Ross1910,Macdonald1957,Anderson1991}  is the basis of most recent work on mathematical modelling of vector-borne infectious diseases like dengue, malaria, and Zika.
Smith et al.~\cite{Smith2012} conducted a historical review highlighting contributions from mathematicians and scientists over 70 years, emphasizing the central role of the Ross--Macdonald model in studying mosquito-borne disease transmission and control strategies. 

As the current manuscript focuses on dengue, we cite two recent reviews~\cite{Aguiar2022,Ogunlade2023}: Aguiar \textit{et al.}~\cite{Aguiar2022} presents a 10-year systematic review on mathematical models for dengue fever epidemiology, centred on models that take into account the multiple strains of dengue and the within-host interactions between viral replication
and immune response. The focus of~\cite{Ogunlade2023} is instead on models of mosquito population dynamics and the control strategies that could limit mosquito density or infection spread.

Extensive biological research has been conducted to understand mosquito behaviour, particularly host-seeking processes, but this aspect has received limited attention in mathematical modelling. 
Indeed, the host-seeking process of mosquitoes entails three stages according to the seeking distance.
In long distance, mosquitoes depend on the odor to seek hosts; in medium distance, they rely on both odors and carbon dioxide (CO$_2$); in short distance, they can seek hosts by more direct approaches, such as vision, heat, body moisture~\cite{Takken1991}. 
As mentioned above, the odor plays a critical role in the whole host-seeking process, so many investigations focusing on mosquito attraction to odors~\cite{Mathew2013,Mcbride2016,Cout2022} have been conducted; consequently, vector management programs should consider mosquito attractants~\cite{Dormont2021} and the differential attractiveness of humans~\cite{Martinez2020}. 
In this respect, Kelly \textit{et al.} found in 2015 that malaria parasites can produce volatiles that can attract mosquitoes~\cite{Kelly2015}.
Similarly, in 2022, Zhang \textit{et al.} found that flavivirus can change the skin microbiota population of the host and produce more mosquito attractants to enhance the host exposure~\cite{Zhang2022}. 

The dynamics of a vector-borne infection is definitely affected by a differential attractiveness to mosquitoes of infected individuals. Indeed, Abboubakar \textit{et al.}~\cite{Abboubakar2016} studied a mathematical model for malaria infection that included the effect of malaria infection on mosquito biting behaviour and attractiveness of humans. Thapa and Ghersi~\cite{Thapa2023} developed an agent-based model for a generic vector-borne infection showing how the preferential attraction of vectors to infected hosts facilitates infection persistence. 

One of the suggestions proposed by Dormont \textit{et al.}~\cite{Dormont2021} was the use of lure-and-kill odor-based traps to control the number of mosquitoes. Here, we present a mathematical model aiming at studying the effect of this strategy on the dynamics of a vector-borne infection. The model is based on the Ross--Macdonald framework, and takes into account also the differential attractiveness to mosquitoes of infected humans, mentioned above. As far as we know, this is the first study of this type.
	
The paper is organized as follows.
Section \ref{sect2} presents the model, showing its well-posedness and the basic dynamical properties.
Section \ref{sect3} finds the conditions for the existence of endemic equilibria and their number.
In Section \ref{sect4}, we present the criterion for backward bifurcation of equilibria at $\mathcal{R}_0 =1$, and prove the local stability of the endemic equilibrium under conditions that allow for model simplification.
Section \ref{sect5} shows numerical simulations illustrating the effect of the parameters and potential infection dynamics.
Finally, we set a brief conclusion and discussion about our model in Section \ref{sect6}.

\section{Odor-baited trap model}  \label{sect2}
Based on the classical Ross--Macdonald-type models proposed by Ross in 1911 and developed by Macdonald in 1957~\cite{Ross1910,Macdonald1957,Anderson1991}, we divide the host into three compartments, including susceptible, infected, and recovered individuals, denoting them by $S_H,\, I_H$, and $R_H$, respectively.
On the other hand, we divide the vectors (mosquitoes) into two compartments, including susceptible and infected mosquitoes, denoting by $S_M$ and $I_M$ the density of susceptible and infected mosquitoes. For the sake of simplicity, we neglect the incubation periods between the moment of infection and the start of infectiousness.

As described in the Introduction, the model also considers the presence of odor-baited traps, whose number is $T$. We assume that the mosquitoes can only be attracted by odor-baited traps and hosts.
The attractive probability of an odor-baited trap, of a susceptible/recovered host, and of an infected host are denoted as $P_T$, $P_S$, and $P_I$, respectively, with $P_I\ge P_S$.
This means that the probability that a mosquito is attracted to a trap is
\[
\frac{P_T T}{P_T T + P_S (S_H + R_H) + P_I I_H},
\]
and similarly for the different types of hosts.

We can divide these expressions by $P_I$ and introduce the two parameters
$\alpha = P_S/P_I \le 1$ (the relative attractiveness of susceptible vs.~infected hosts) and 
$L = P_T T/P_I$, the effective number of traps, measured in the attractive power of an equivalent number of infected hosts.

It is also useful to introduce the quantity
\[ A(t) = L+\alpha(S_H(t)+R_H(t))+I_H(t).\]
Then, we can write the probability for a mosquito of being attracted to a trap, to a susceptible, infected, or recovered host as, respectively,
\begin{gather*}
  \frac{L}{A(t)}, \qquad \frac{\alpha S_H}{A(t)}, \qquad \frac{I_H}{A(t)}, \qquad \frac{\alpha R_H}{A(t)}. 
  \end{gather*}
Using these expressions, we can compute the rate at which susceptible hosts (or mosquitoes) become infected.
Let $b$ be the biting rate of mosquitoes, $P_{mh}$ ($P_{hm}$) the probability that a susceptible host (mosquito) gets infected when bitten by (biting) an infected mosquito (host). 

Then the rates at which the hosts or the mosquitoes get infected are, respectively, 
$$P_{mh} b I_M \frac{\alpha S_H}{A}\quad\mbox{and}\quad P_{hm}b S_M \frac{ I_H}{A}.$$	
For writing convenience, we set
    \begin{equation}
        \label{beta_lambda}
        \beta_H = P_{mh} b,\qquad \beta_M =P_{hm} b, \qquad \lambda_h(I_M,A) = \beta_H \alpha  \frac{I_M}{A}, \qquad \lambda_m(I_H,A) = \beta_M  \frac{I_H}{A},
    \end{equation}
where $\lambda_h$ and $\lambda_m$ are the force of infection for susceptible hosts and mosquitoes, respectively. Furthermore, we assume that mosquitoes that get attracted to traps die with probability $\rho$. Then the mortality caused by the traps is $\rho b L/A$.

The other parameters  of the model represent the demography and the course of infection in hosts. Precisely,
$\Lambda_H$ and $ \Lambda_M$ represent the constant recruitment of host and adult mosquitoes, respectively, $d$ and $\mu$ are the natural death rates of hosts and mosquitoes, respectively, $\gamma$ is the recovery rate of hosts, $\xi$ is the additional mortality of hosts caused by disease and $\delta$ is the decay rate of adaptive immunity.

Summarizing, the host-vector model with odor-baited trap writes
\begin{equation} \label{endemic}
    \left\{
    \begin{aligned}
        S_H' &= \Lambda_H - d S_H - \beta_H \alpha  \frac{I_M}{A} S_H+\delta R_H,\\
        I_H' &= \beta_H \alpha  \frac{I_M}{A} S_H - (d + \gamma + \xi)I_H,\\
        R_H' &= \gamma I_H -\delta R_H - d R_H,\\
        S_M' &= \Lambda_M -\mu S_M - \beta_M   \frac{I_H}{A} S_M - b \rho S_M \frac{L}{A},\\
        I_M' &=  \beta_M   \frac{I_H}{A} S_M -\mu I_M - b \rho I_M \frac{L}{A}.
    \end{aligned}
    \right.
\end{equation}

It is actually convenient to introduce, instead of $\Lambda_H$ and $\Lambda_M$, the population equilibrium values in the absence of infections and traps. In that case, the total host population $N_H = S_H+I_H + R_H$ satisfies the equation 
$$ N'_H = \Lambda_H - d N_H,$$
whose equilibrium value is $\bar N_H = \Lambda_H/d$. In the absence of traps, $N_M = S_M+I_M $ satisfies the equation 
$$ N'_M = \Lambda_M - \mu N_M,$$ whose equilibrium value is $\bar N_M = \Lambda_M/\mu$.\\
The equations for $S_H$ and $S_M$ in~\eqref{endemic} can then be rewritten as 
\begin{equation} \label{endemic2}
    \left\{
    \begin{aligned}
        S_H' &= d(\bar N_H - S_H) - \beta_H \alpha  \frac{I_M}{A} S_H+\delta R_H,\\
        S_M' &= \mu(\bar N_M - S_M) - \beta_M   \frac{I_H}{A} S_M - b \rho S_M \frac{L}{A}.
    \end{aligned}
    \right.
\end{equation} 


\subsection{Preliminaries}
Here we show that System~\eqref{endemic} is well-posed and calculate the basic reproduction number $\mathcal{R}_0$ from which one obtains the condition for the local stability of the disease-free equilibrium (DFE).

\medskip
\begin{proposition}
    The set 
    \[
    \mathcal{D} := \left\{x=(S_H,I_H,R_H,S_M,I_M)\in \mathbb{R}_+^5:\,\ S_H+I_H+R_H \le \bar N_H,\ S_M+I_M \le \bar N_M \right\}
    \]
    is positively invariant and attracting all initial values in $\R_+^5$.
\end{proposition}
\begin{proof}
    To show that all variables remain nonnegative, it suffices to observe that, if one of them (say $S_M$) is equal to 0, its derivative is nonnegative.

Regarding the other conditions, the total host $H(t)= S_H(t)+I_H(t)+ R_H(t)$ satisfies 
    $$ 
    H'(t) = d (\bar N_H - H(t)) - \xi I_H(t) \le d (\bar N_H - H(t)),
    $$
from which it follows that $H(t_0) \le \bar N_H$ implies $H(t) \le \bar N_H$ for $t \ge t_0$, and that, if $H(t) > \bar N_H$ $\forall\ t \ge 0$, then $H'(t) \le 0$ and $H(t) \to \bar N_H$ as $t \to +\infty$.

A similar argument holds for $M(t) = S_M(t) + I_M(t)$.   
\end{proof}

\begin{figure}
    \centering
    \includegraphics[width=0.5\linewidth]{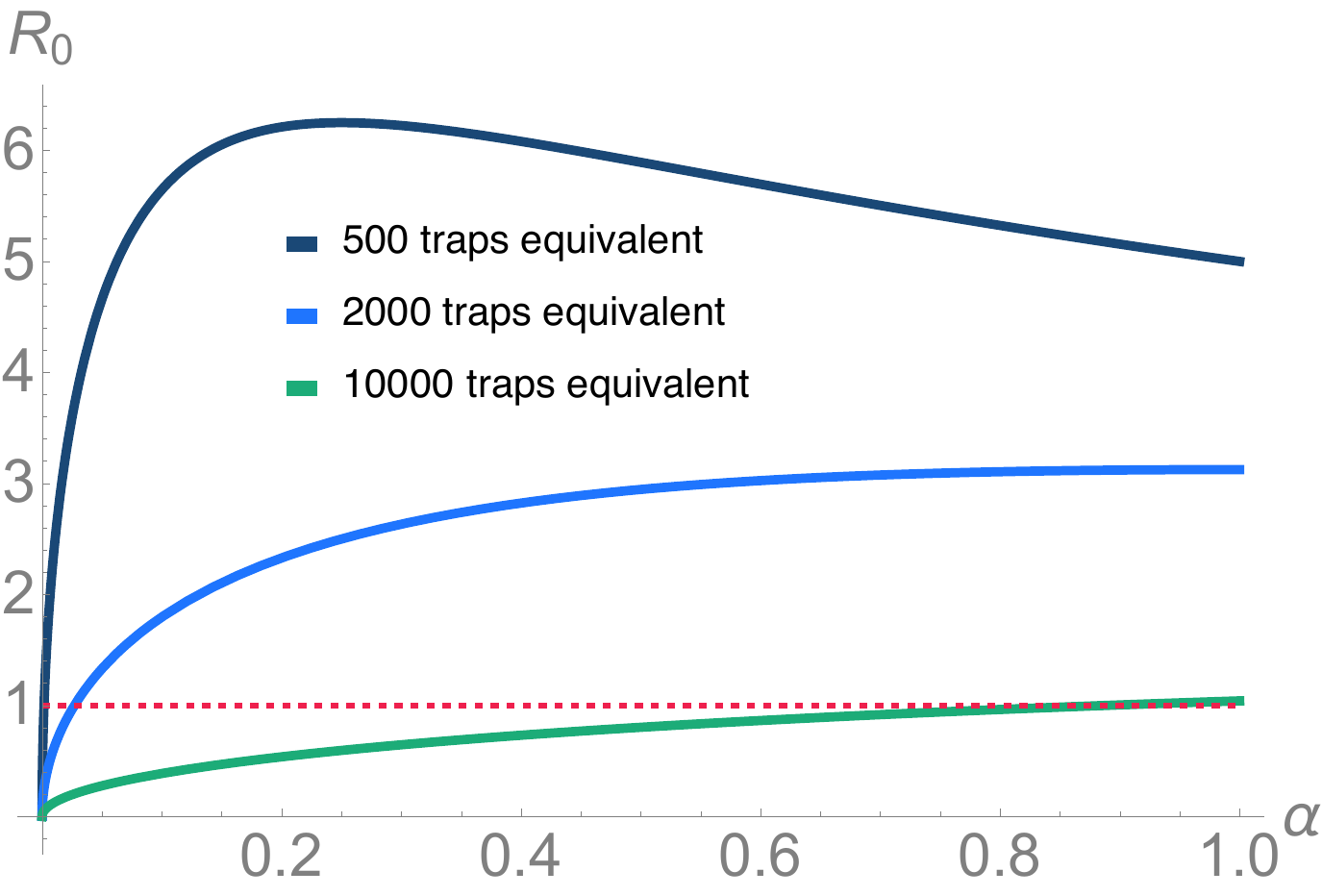}
     \captionsetup{width=14cm}
    \captionsetup{font=small}
    \caption{$\mathcal{R}_0$ as a function of $\alpha$ for $L$ equal to 500, 2000 or 10000. Other parameter values are as in Table \ref{Tab1}.}
    \label{fig_R0}
\end{figure}

We next find the basic reproduction number $\mathcal{R}_0$ by the method of next generation matrix~\cite{Driessche2002}.
First, letting $I_H = 0, R_H =0, I_M = 0$, the DFE of System~\eqref{endemic} is $E_0 := (\bar N_H,0,0,S_{M0},0)$, where  
\begin{equation}
    \label{DFE}
    S_{M0} = \bar N_M \frac{\mu}{\mu + \frac{b \rho L}{A_{0}}},  \quad\mbox{with }\  A_{0}= L +\alpha \bar N_H.
\end{equation}
Now, we can write 
$$\begin{pmatrix}
    I'_H \\ I'_M
\end{pmatrix} = \mathcal{F} - \mathcal{V}, \quad\mbox{with}\quad
\mathcal{F} = \begin{pmatrix}
 \beta_H I_M \frac{\alpha S_H}{A}\\[0.5em]
  \beta_M S_M \frac{ I_H}{A}
\end{pmatrix},\quad 
\mathcal{V} = \begin{pmatrix}
(\gamma + d + \xi) I_H\\[0.5em]
\mu I_M + b \rho I_M \frac{L}{A}
\end{pmatrix},
$$
so that the Jacobian matrix of $\mathcal{F}$ and $\mathcal{V}$ at DFE are respectively
$$
F = D\mathcal{F}= 
    \begin{pmatrix}
		0& \frac{\beta_H \alpha \bar N_H}{A_{0}}\\
		\frac{\beta_M  S_{M0}}{A_{0}}&0
    \end{pmatrix},\quad 
    V = D\mathcal{V}= 
    \begin{pmatrix}
	\gamma + d + \xi & 0\\
	0& \mu + \frac{b \rho L}{A_{0}}
    \end{pmatrix}.
$$
Thus, from  $\mathcal{R}_0 = \rho(F\cdot V^{-1}),$ we can obtain 
\begin{equation}\label{R_0}
    \mathcal{R}_0 = \sqrt{\frac{\beta_H\beta_M\alpha \bar N_H S_{M0}}{A_{0}^2(\gamma+d+\xi)(\mu + \frac{b \rho L}{A_{0} })}}=\sqrt{\frac{\beta_H\beta_M\bar N_H\bar N_M \alpha \mu}{(\gamma+d+\xi)[(\mu + b \rho )L+\alpha  \mu \bar N_H]^2}}.
\end{equation}
Notice that $\Ro$ depends only on the ratios $\bar N_M/\bar N_H$ (vector to host) and $L/\bar N_H$ (traps to hosts), since
\[ \mathcal{R}_0 = \sqrt{\frac{\beta_H\beta_M\frac{\bar N_M}{\bar N_H} \alpha }{(\gamma+d+\xi)\mu[(1 + \frac{b \rho}{\mu} )\frac{L}{\bar N_H}+\alpha   ]^2}}.\]
From this expression, one sees that, as a function of $\alpha$ (the relative attractiveness of susceptibles relative to infected), $\Ro$ is initially increasing (if $L>0$) and then decreasing, as shown in Figure~\ref{fig_R0} for different values of the traps equivalence $L$.

It is also easy to check that $F$ and $V$ meet the five assumptions of next generation matrix theory proposed by Driessche~\cite{Driessche2002}.
Thus, we can directly have the following theorem about the local stability of $E_0$.
\begin{theorem}
    The DFE $E_0$ of System~\eqref{endemic} is locally asymptotically stable as $\mathcal{R}_0 <1$. 
\end{theorem}

\section{Endemic equilibria}\label{sect3}
In this section, we will study the conditions for the existence of endemic equilibria of System~\eqref{endemic} and their number, investigating the conditions for the existence of two endemic equilibria (EE) when $\mathcal{R}_0<1$. 

\subsection{Computation of endemic equilibria}
We now investigate the EE of System~\eqref{endemic} and find the condition of its existence. Supposing System~\eqref{endemic} has a positive root, denoted as $E^* = [S_H^*,I_H^*,R_H^*,S_M^*,I_M^*]$, then we can denote the infection force function at balance
\begin{equation}\label{lambda_h_v}
    \lambda_h^* = \frac{\beta_H \alpha I_M^*}{A^*},\qquad \lambda_m^* = \frac{\beta_M I_H^*}{A^*}.
\end{equation}
Then we use $\lambda^*_h$ to express other variables, and they are
\begin{equation}\label{SIRSI}
    \begin{aligned}
        R_H^* &= \frac{\gamma}{\delta + d}I_H^*,\qquad I_H^* = \frac{\lambda_h^*}{k_1} S_H^*, \qquad S_H^* = \frac{\Lambda_H}{d + k_2\lambda_h^*},\\[1mm]
        S_M^* &= \frac{\Lambda_M}{\mu+\lambda_m^*+b \rho L /A^*},\qquad I_M^* = \frac{\lambda_m^*}{\mu+b \rho L/A^*}S_M^*,\\[2mm]
        A^* &= L + \alpha(S_H^* + R_H^*) +  I_H^*,\\[1mm]
        k_1 &= \gamma+ d + \xi,\qquad k_2 = \frac{d k_1+\delta(d+\xi)}{(d + \delta)k_1}.
    \end{aligned}
\end{equation}
Now, we need to find $\lambda_h^*$ and then, based on equation~\eqref{lambda_h_v} and~\eqref{SIRSI}, we can get the EE $E^*$.
	
First, taking $S_M^*$ and $I_M^*$ in~\eqref{SIRSI} into $\lambda_h^*$ in~\eqref{lambda_h_v}, we have 
\begin{equation}\nonumber
    \lambda_h^* = \frac{\beta_H \alpha \Lambda_M}{\mu A^* + b\rho L} \frac{\lambda_m^* A^*}{(\mu +\lambda_m^*)A^* + b \rho L}.
\end{equation}
Since $\lambda_m^* A^* = \beta_M  I_H^*$ in~\eqref{lambda_h_v}, we next have after eliminating the denominator
\begin{equation} \label{d+k2lambda}
    (\mu A^* + b\rho L)(\mu A^* + b\rho L + \beta_M  I_H^*)	\lambda_h^* = \beta_H \alpha \Lambda_M \beta_M  I_H^*.
\end{equation}
From $I_H^*$, $S_H^*$ and $A^*$ in~\eqref{lambda_h_v}, we have $I_H^*= \Lambda_H \lambda_h^*/[k_1(d+k_2 \lambda_h^*)]$ and
\begin{equation}\nonumber
    \begin{aligned}
        &(\mu A^*+b \rho L)(d+k_2\lambda_h^*) \\
        =& b\rho L (d+k_2\lambda_h^*) + \mu A^*(d+k_2\lambda_h^*)\\
        =& (b\rho L+\mu L)(d+k_2\lambda_h^*)+\mu\Lambda_H\left[\alpha + \left(\frac{\gamma}{d+\delta} \alpha+1\right)\frac{\lambda_h^*}{k_1}\right]\\
        \triangleq& a_0 + b_0 \lambda_h^*,
    \end{aligned}
\end{equation}	
where 
$$ 
a_ 0 = \mu (L d + \alpha \Lambda_H) +b\rho Ld, \quad \text{and} \quad  b_0 = \mu \left[L k_2 +\frac{\Lambda_H}{k_1}\left( \frac{\gamma}{d +\delta}\alpha + 1\right)\right] + b \rho L k_2.
$$
Thus, equation~\eqref{d+k2lambda} can be transformed by multiplying $(d+k_2\lambda^*_h)^2$ on both sides
\begin{equation}\nonumber
    \beta_H \alpha \Lambda_M \beta_M  \Lambda_H \frac{\lambda_h^*}{k_1}(d+k_2\lambda_h^*) =\lambda^*_h(a_0 + b_0 \lambda_h^*)\left[a_0 + \left(b_0 + \frac{\beta_M  \Lambda_H}{k_1}\right)\lambda_h^*\right].
\end{equation} 
Let $Q = \beta_H\Lambda_H\beta_M \Lambda_M,$ and after some algebraic operations, we finally have
\begin{equation}\nonumber
    \lambda_h^* (A {\lambda_h^*}^2 + B \lambda_h^* +C)= 0,
\end{equation} 
where 
\begin{equation}\nonumber
    A = b_0 (b_0 k_1 +\beta_M  \Lambda_H), B =2 a_0 b_0 k_1 + a_0 \beta_M  \Lambda_H - Q k_2 \alpha  , C = a_0^2 k_1- Q \alpha  d.
\end{equation}
Specifically, if $\lambda_h^* = 0$, we get the DFE $E_0$.
If not, we need to investigate the equation
\begin{equation}\label{rooteq}
    A {\lambda_h^*}^2 + B \lambda_h^* + C= 0.
\end{equation}
Noticing $A>0$, thus the number of roots in~\eqref{rooteq} depends on $B$ and $C$.
Particularly, we can rewrite $B$ and $C$ as 
$$
B = \frac{a_0^2 k_1k_2}{d}(R^2-\mathcal{R}_0^2), \quad C= a_0^2 k_1 (1-\mathcal{R}_0^2),
$$
where $\mathcal{R}_0$ can be rewritten, from~\eqref{R_0}, as follows
$$
\Ro = \sqrt{\dfrac{Q \alpha  d}{k_1 a_0^2}},
$$
and we can define
$$ 
R = \sqrt{\frac{d(2b_0k_1 + \beta_M  \Lambda_H)}{a_0 k_1  k_2}}. 
$$
According to its coefficients, the number of positive equilibria to equation~\eqref{rooteq} can be summarized as follows
\begin{itemize}
    \item (1): equation~\eqref{rooteq} has a unique positive root as $C<0$, i.e. $\mathcal{R}_0 >1$;
    \item (2): equation~\eqref{rooteq} has a unique positive root as 
            \begin{equation} \label{one_root}
                (C = 0, \;B<0)\quad {\rm or}\quad (C>0,\;B<0,\; B^2-4AC =0); 
            \end{equation}	
    \item (3): equation~\eqref{rooteq} has two positive roots as 
            \begin{equation}\label{2_root}
                B<0,\; C>0 \, (\mathcal{R}_0<1) \quad {\rm and} \quad  B^2-4AC >0;
            \end{equation}
    \item (4): equation~\eqref{rooteq} has no positive roots otherwise.	
\end{itemize}
Finally, following the number of roots to equation~\eqref{2_root}, the number of endemic equilibria of System~\eqref{endemic} can be concluded in the following theorem.
\begin{theorem}
    System~\eqref{endemic} has a unique positive equilibrium if $\mathcal{R}_0 >1$ or the condition~\eqref{one_root} holds, two positive equilibria if the condition~\eqref{2_root} holds, and no positive equilibrium otherwise.
\end{theorem}


\subsection{Backward bifurcation} \label{sec32}

As shown by the previous analyses of positive roots to equation~\eqref{rooteq}, System~\eqref{endemic} will have two positive equilibria under condition~\eqref{2_root}, which implies~$\mathcal{R}_0<1$. This result is an indication of a backward bifurcation of the DFE~$E_0$ at~$\mathcal{R}_0 = 1$. 

Since condition~\eqref{2_root} is not easy to interpret, we then study the conditions under which backward bifurcation occurs,  following the method proposed by Boldin~\cite[Theorem 2.2]{Boldin2006}.

There it is assumed that a system~$ x' = f(x,\tau)$, where~$\tau$ is the bifurcation parameter, can be written as 
\begin{equation}\label{eq_yz}
    x = 
    \begin{pmatrix}
        y \\ z
    \end{pmatrix}, \qquad
    \left \{
    \begin{aligned}
        y' &= G(y,z,\tau) y,\\
        z' &= h(y,z,\tau), 
    \end{aligned}
    \right. 
\end{equation}
where $y \in \R^m$, $z \in \R^n$, $G$ is (for all values of $y$, $z$ and $\tau$) an $m \times m$ matrix.

It is further assumed that, for all $\tau$, there exists an equilibrium $e(\tau)=(0,z_0(\tau))$. The stability of $e(\tau)$ can be studied through the Jacobian matrix
$$
f_x(e(\tau)) = 
\begin{pmatrix}
    G(e(\tau),\tau) & 0 \\[1mm] h_y(e(\tau),\tau)&h_z(e(\tau),\tau)
\end{pmatrix}. 
$$
It then follows that the stability can be ascertained by studying the eigenvalues of $G(e(\tau),\tau)$ and of $h_z(e(\tau),\tau)$. We first assume that the spectral bound $s$ of $h_z$ satisfies
$$
s(h_z(e(\tau),\tau)) < 0 \quad \forall \ \tau,
$$
so that the stability of $e(\tau)$ depends only on the matrix $G(e(\tau),\tau)$.

We assume that $G(e(\tau),\tau)$ has the typical structure of epidemic models 
$$ 
G(e(\tau), \tau) = F(\tau)-V(\tau),
$$
with $F(\tau)$ and $V^{-1}(\tau)$ non-negative matrices so that defining 
$$
\Ro(\tau)=\rho(F(\tau)V^{-1}(\tau)),\qquad s(G(e(\tau)),\tau) \ge 0 \iff \Ro(\tau) \geq 1.
$$

We then assume that $\Ro(\tau)$ crosses $1$ (with positive speed) at $\tau = 0$, meaning that $e(\tau)$ will be asymptotically stable for $\tau < 0$ and unstable for $\tau >0$.

System~\eqref{endemic} satisfies all these assumptions by setting
$$
y = 
\begin{pmatrix}
    I_H \\ I_M
\end{pmatrix}
,\quad z = 
\begin{pmatrix}
    S_H \\ R_H \\ S_M
\end{pmatrix},
$$
choosing any model parameter as $\tau$ (shifted so as to have $\Ro=1$ at $\tau =0$), and obtaining $e(\tau)=E_0$.

To find the structure of the bifurcation at $\tau =0$, Boldin studies the center manifold of the system at $\tau = 0$ ($\Ro(\tau) = 1$). Let then $H_z = h_z(E_0,0)$ and $H_y = h_y(E_0,0)$, $w$ and $v$ are the left and right eigenvectors of $G_0=G(E_0,0)$ corresponding to the eigenvalue $0$, normalized so that $v\cdot w = 1$. System~\eqref{eq_yz} exhibits a forward (with existence of a stable positive equilibrium close to $E_0$ for $\tau > 0$) at $\tau=1$ or a backward bifurcation (with an unstable positive equilibrium close to $E_0$ for $\tau < 0$) depending on whether $M <0$ or $M >0$, where
\begin{equation} \label{M1M2}
    \begin{aligned}
        M &= M_1 - 2 M_2,\qquad{\rm and}\\
        M_1 &= \sum_{i,j,k = 1,\cdots,m} w_i \left(\frac{\partial  G_{ij}(E_0)}{\partial y_k} + \frac{\partial  G_{ik}(E_0)}{\partial y_j}\right) v_j  v_k,\\
        M_2 &= \sum_{\substack{i,j=1,\dots,m \\ k=1,\dots,n}} w_i \frac{\partial G_{ij}(E_0)}{\partial z_k} v_j (H_z^{-1}H_y v)_k.
    \end{aligned}
\end{equation}

Performing all the computations (reported in Appendix~\ref{app1}), we obtain the following theorem.
\begin{theorem}\label{bifur}
    System~\eqref{endemic} has a backward bifurcation at DFE $E_0$ at $\mathcal{R}_0=1$ if and only if 
    \begin{equation}\label{cond_back}
    \frac{L}{\bar N_H} < \sqrt{\frac{\mu}{\mu+b\rho}}\alpha,
\end{equation}
and
\begin{equation}
    \label{cond1_back}
\frac{\xi}{d}\left( \frac{\alpha(2 \mu A_{0} + b \rho L)}{A_{0}(\mu A_{0} + b \rho L)} - \frac{1}{\bar N_{H}}\right) > C,
\end{equation}
where
 \begin{equation}\label{C_main}
    C = \frac{( 1- \alpha)(2 \mu A_{0} + b \rho L)}{A_{0}(\mu A_{0} + b \rho L)} + \frac{ \beta_M}{\mu A_{0} + b \rho L} + \frac{1}{\bar N_{H}}\left(1 + \frac{\gamma}{d+\delta}\right) .
\end{equation}
\end{theorem}
\begin{remark}
Note that $C > 0$, since $\alpha < 1$ and remember that $A_0$ has been defined in~\eqref{DFE}.

Condition~\eqref{cond_back} (that involves only a few parameters of the model) implies that backward bifurcation is possible only if $L/\bar N_H$ is not too large; increasing the lethality $\rho$ of traps also makes backward bifurcation less likely.

If~\eqref{cond_back} is satisfied, then backward bifurcation occurs only if the ratio $\xi/d$ between mortality induced by the infection and natural mortality is large enough.
\end{remark}

\section{Stability of the endemic equilibrium} \label{sect4}
We study here the stability of the endemic equilibrium of System~\eqref{endemic} when the total population of hosts and mosquitoes is constant and not influenced by the infection. This requires that the extra mortality of the host caused by disease and the mosquito death caused by odor-baited trap may be neglected, i.e., we assume $\xi = \rho =0$. 

From the previous Section, we already know that backward bifurcation is impossible under these assumptions.
Thus, System~\eqref{endemic} has a unique DFE and no endemic equilibria when $\mathcal{R}_0<1$.
We then need to study the local stability of the endemic equilibrium, only under the scenario of a unique one for $\mathcal{R}_0>1$.
    
The total population of hosts $(S_H+I_H+R_H)$ and mosquitoes $(S_M+I_M)$ is asymptotically stable as $\xi = \rho = 0$, namely
\begin{equation}\label{total_popu}
    \lim_{t\to+\infty}S_H+I_H+R_H = N_H := \frac{\Lambda_H}{d},\quad \text{and} \quad  \lim_{t\to +\infty} S_M+I_M = N_M := \frac{\Lambda_M}{\mu}.
\end{equation}
Combining it with the result in equation~\eqref{total_popu}, we can reduce System~\eqref{endemic} to a system in the three variables~$S_H,\,I_H,\,S_M$
\begin{equation}\label{reduce_model}
    \left\{
    \begin{aligned}
        S_H' &= \Lambda_H - d S_H - \beta_H \alpha \frac{N_M-S_M}{A} S_H+\delta (N_H-S_H-I_H),\\
        I_H' &= \beta_H \alpha \frac{N_M-S_M}{A}S_H - (d + \gamma)I_H,\\
        S_M' &= \Lambda_M  -\mu S_M - \beta_M  \frac{I_H}{A}  S_M,
    \end{aligned}
    \right.
\end{equation}
with
$$
A = \alpha N_H +(1-\alpha) I_H + L.
$$
If we show that the endemic equilibrium of System~\eqref{reduce_model} is locally asymptotically stable, then the endemic equilibrium of System~\eqref{endemic} is locally asymptotically stable because of the theorems on asymptotically autonomous equations.

Let, for $\mathcal{R}_0>1$, the endemic equilibrium of System~\eqref{reduce_model} be denoted by~$\tilde{E}^* = (S_H^*,I_H^*,S_M^*)$. 
The Jacobian matrix of~\eqref{reduce_model} at $\tilde{E}^*$ is 
\begin{equation}\nonumber
    J = 
    \begin{pmatrix}
        -d -\beta_H I_M^* \frac{\alpha}{A^*}-\delta & \beta_H I_M^*  S_H^* \frac{\alpha}{{A^*}^2} (1-\alpha)-\delta & \beta_H \frac{\alpha S_H^*}{A^*}\\[0.3em]
        \beta_H I_M^* \frac{\alpha}{A^*} & - \beta_H I_M^*  S_H^* \frac{\alpha}{{A^*}^2} (1-\alpha) - (\gamma +d) & -\beta_H \frac{\alpha S_H^*}{A^*}\\[0.3em]
        0&-\beta_M \frac{S_M^* }{A^*}\left(1- \frac{(1-\alpha)\andrea{I_H^*}}{A^*}\right)& -\mu -\beta_M  \frac{I_H^*}{A^*} 
    \end{pmatrix}.
\end{equation}
By using the values of the variables at the equilibrium, we have
\begin{equation}\nonumber
    J= 
    \begin{pmatrix}
        -d - (\gamma + d)\frac{I^*_H}{S_H^*}-\delta & \frac{I^*_H}{A^*}(\gamma+d)(1-\alpha) - \delta       & (\gamma+d)\frac{I^*_H}{I^*_M}\\[0.3em]
        (\gamma+d)\frac{I^*_H}{S^*_H}               & - \frac{I^*_H}{A^*}(\gamma+d)(1-\alpha) - (\gamma+d) & - (\gamma+d)\frac{I^*_H}{I^*_M}\\[0.3em]
        0                                           & -\mu \frac{I^*_M}{I^*_H} \left(1- \frac{(1-\alpha)\andrea{I_H^*}}{A^*}\right) &  - \mu\frac{N_M}{S^*_M}
    \end{pmatrix}.
\end{equation}
For writing convenience, we denote 
\begin{equation}\nonumber
    \begin{aligned}
        Q &= (\gamma + d)\frac{I^*_H}{S_H^*}, \quad P =\frac{I^*_H}{A^*}(\gamma+d)(1-\alpha),\\
        S &=(\gamma+d)\frac{I^*_H}{I^*_M}, \quad  N = \mu \frac{I^*_M}{I^*_H} (1- \frac{(1-\alpha)\andrea{I_H^*}}{A^*}), \quad U = \mu\frac{N_M}{S^*_M},
    \end{aligned}
\end{equation}
which are all positive quantities.

Thanks to the Routh--Hurwitz criterion~\cite{Dejesus1987,Prasolov2009}, all eigenvalues of the $3 \times 3$ matrix $J$ have a negative real part if and only if
$$ 
a_1 > 0,\quad a_2 > 0, \quad a_3 > 0 \quad \mbox{ and } \quad a_1 a_2 - a_3 >0,
$$
where
$a_1 =  -{\rm tr}(J)$, $a_3 =-{\rm det}(J)$, and $a_2$ is the sum of the principal minors.

We get
\[\begin{split}
    a_1 &= P+Q+2d + \delta + \gamma + U > 0,\\
    a_2 &= (d+Q+\delta) (P + \gamma+d)+(d+Q+\delta)U\\
            &+(P+\gamma+d)U-NS+Q(-P+\delta).
\end{split}\]
Exploiting the identity
$$
(\gamma +d)U - NS = (\gamma+d)\mu\left(\frac{I_M^*}{S^*_M} + \frac{(1-\alpha)\andrea{I_H^*}}{A^*}\right),
$$
one arrives at
\[\begin{split} 
    a_2 &= P(d+\delta)+(d+\gamma)\left(Q +d+\delta+U+\mu\left(\andrea{\frac{I_M^*}{S^*_M} + \frac{(1-\alpha)I_H^*}{A^*}}\right)\right)\\
    &+U(P+Q+d + \delta)+Q\delta>0.
\end{split}\]
Furthermore
\[\begin{split} 
    a_3 &= U(P+\gamma+d)(d+\delta+Q) -NS(d+\delta)-QU(P-\delta)\\
    &= [Q(\gamma+d+\delta)+P(d+\delta)]U+(d+\delta)(\gamma + d)\mu\left(\andrea{\frac{I_M^*}{S^*_M} + \frac{(1-\alpha)I_H^*}{A^*}}\right)>0.
\end{split}\]
Finally
\begin{equation}\nonumber
    \begin{aligned}
        a_1 a_2 - a_3 &= \left[(d+\gamma)(d+\delta+U)+U(P+Q+d+\delta)+Q\delta \right] a_1 \\
        &+[P(d+\delta)+Q(d+\gamma)](P+Q+2d + \delta + \gamma)\\
        &+(d+\gamma)\mu\left(\andrea{\frac{I_M^*}{S^*_M} + \frac{(1-\alpha)I_H^*}{A^*}}\right) >0.
    \end{aligned}
\end{equation}

According to the Routh--Hurwitz criterion, it turns out that all eigenvalues of the matrix $J$ have negative real parts.
Therefore, we know that the endemic equilibrium~$\tilde{E}^*$ of System~\eqref{reduce_model} is locally asymptotically stable, and thus the following stability result holds naturally.
\begin{theorem}
    The endemic equilibrium $E^*$ of System~\eqref{endemic} is locally and asymptotically stable if $\mathcal{R}_0>1$ and $\xi = 0,\; \rho =0$.
\end{theorem}
\begin{remark}
If beyond $\xi=\rho=0$, one assumes $L=\delta=0$ and $\alpha=1$, it is known that the endemic equilibrium is asymptotically stable when $\Ro > 1$, since in that case System~\eqref{endemic} is a special case of the SEIR-SI model with delay studied by Cai \textit{et al.}~\cite{Cai2017}, who proved the global asymptotic stability through a Lyapunov function. The same conclusion could potentially be obtained within the current model through a suitable adjustment of that Lyapunov function.
\end{remark}

\section{Numerical simulations} \label{sect5}
In this section, the theoretical results will be verified, and the influence of odor-baited traps on $\mathcal{R}_0$ will be investigated.

In this section, dengue is taken as a representative example of a mosquito-borne disease for the implementation of numerical simulations. The parameters of System~\eqref{endemic} are taken from the existing literature and listed in Table~\ref{Tab1}. We set the population of humans and mosquitoes as $10^4$ and $10^5$, respectively, but the relevant parameter is the proportion of mosquitoes to hosts, set as $10:1$ unless otherwise stated (namely, in panel d) of Figure \ref{fig_EE_vs_pars2}).
\begin{table}[H]
    \centering
    \begin{tabular}{cccc}
        \toprule
        Parameters &Description&Base value/range&Reference\\
        \hline
        $\Lambda_H$&Human input rate&$10^4/68.5$ humans/year& assumed\\[1mm]
        $d$ &Human death rate&$1/68.5$ /year&~\cite{Abboubakar2016,Chitnis2008,WHOlifespan,Newton1992}\\[1mm]
        $\xi$ & Human death rate from disease&  $7/10^5$ /day&~\cite{WHOb}$^*$\\[2mm]
        $\gamma$& Human recovery rate& $1/5$ /day&~\cite{Newton1992}\\[2mm]
        $\delta$& Immunity decay rate& $1/1000$ /day& assumed\\[2mm]
        $\Lambda_M$&Mosquito input rate&$10^5/9 $ mosq./day&assumed\\[2mm]
        $\mu$ &Mosquito death rate&$1/10$ /day&\cite{Alphey2011}\\[2mm]
        $b$&Bites per mosquito per day&$1/2$ /day&~\cite{Abboubakar2016,Chitnis2008}\\[2mm]
        $P_{mh}$& \makecell{Probability of disease\\ transmission to human}&$0.75$&\cite{Newton1992} \\[5mm]
        $P_{hm}$& \makecell{Probability of disease\\ transmission to mosquito}&$0.375$&~\cite{Newton1992}$^{**}$\\[5mm]
        $L$&{Total attractiveness of odor-baited traps}&$\geq 0$&--\\[3mm]
        $\alpha$&\makecell{Relative attractive rate of $S_H\;R_H$\\ as compared to infected}&0--1&--\\[5mm]
        $\rho$&Lethality of traps to mosquitoes& 0--1&--\\[1mm]
        \bottomrule
    \end{tabular}\\
    \raggedright \small
    $^*$ Based on the proportion of deaths to reported cases ($7.5\cdot 10^{-4}$), and on the estimate that 1 out of 4 infections gives clinical symptoms and is reported.\\
    $^{**}$ Taking into account the probability of mosquito death during the incubation period, which is not present in the current model.
    \caption{Description and base values/range of parameters appearing in System~\eqref{endemic} used in the numerical simulations (unless otherwise stated).}\label{Tab1}
\end{table}

\subsection{Impact of parameters}

We first consider the influence on $\mathcal{R}_0$ of $L$, $\alpha$, and $\rho$, the parameters left free in Table~\ref{Tab1}.
The expression~\eqref{R_0} reveals that $\mathcal{R}_0$ decreases as $L$ or $\rho$ increases, while the dependence on $\alpha$ is non-monotone, initially increasing and decreasing after the peak at 
$$
\alpha = \left( 1+\frac{b\rho}{\mu} \right)\frac{L}{\bar N_{H}}.
$$ 
Specifically, if the traps are harmless to mosquitoes ($\rho=0$), the peak condition can be simplified as $\alpha \bar N_{H} = L$, that is to say, $\mathcal{R}_0$ takes the maximum value when the attractiveness of the whole susceptible population is equal to that provided by the traps. If $\rho > 0$, the peak is reached when the attractiveness of the whole susceptible population is somewhat higher.

\begin{figure}[H]
    \centering
    \captionsetup{width=14cm}
    \captionsetup{font=small}
    \includegraphics[width=0.48\linewidth]{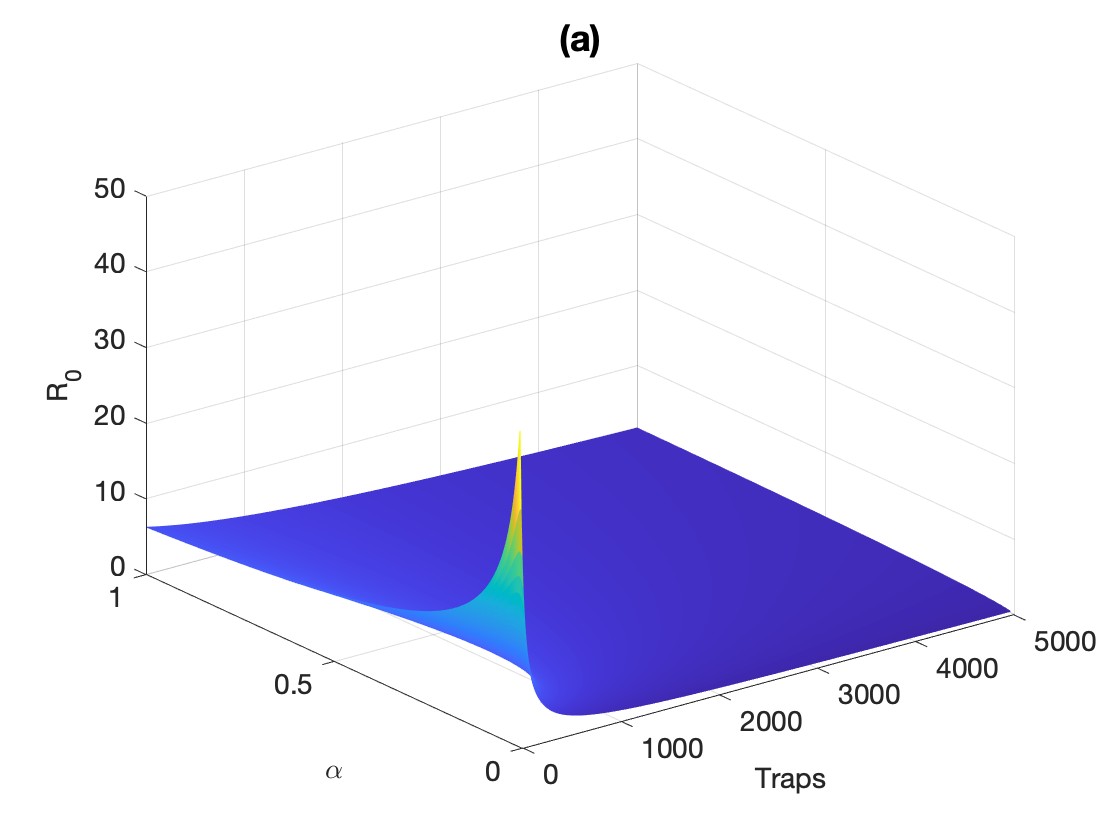}
    \includegraphics[width=0.48\linewidth]{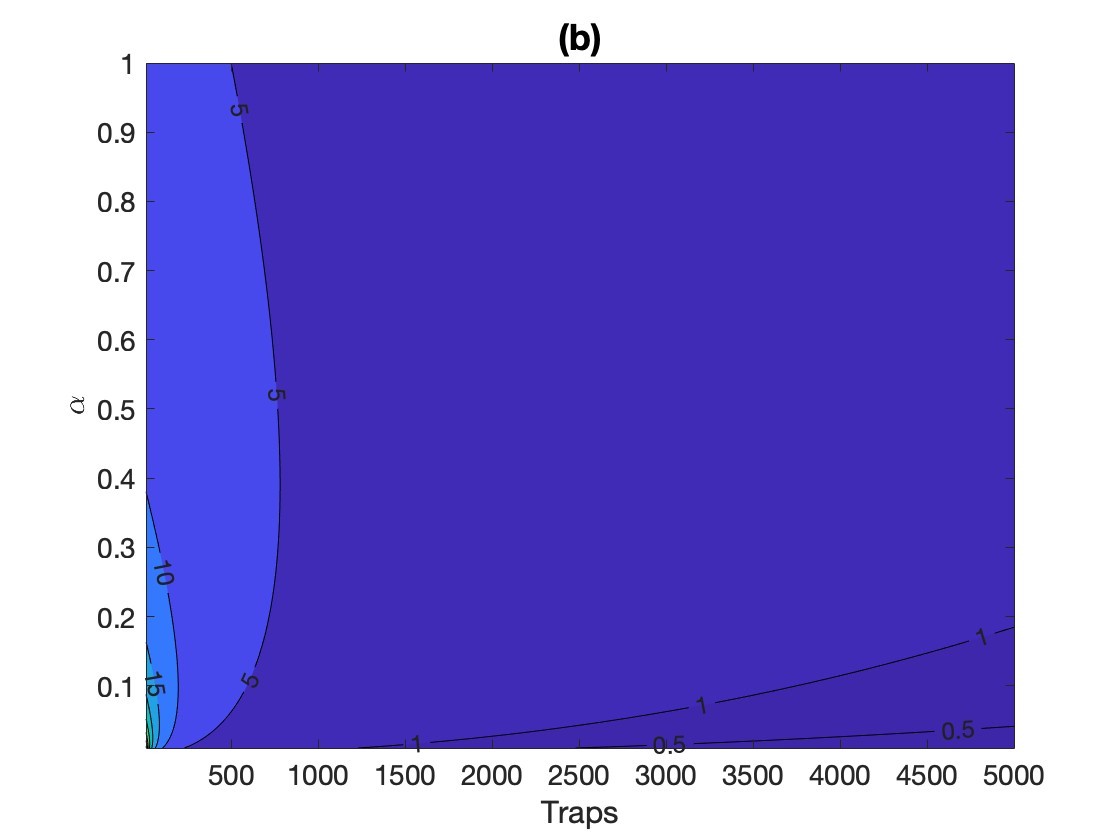}
    \caption{Local surface of $\mathcal{R}_0$ over plane $L\times \alpha$. The total attractiveness $L$ of traps ranges from $5$ to $5000$, the relative attractiveness of susceptible/recovered as compared to infected hosts~$\alpha$ ranges from $0$ to $1$, and the lethality of traps $\rho =0.8$. The other parameters are as in Table~\ref{Tab1}.}
    \label{R0_L_alpha1}
\end{figure}

We illustrate the quantitative effect of $L$ and $\alpha$ on $\mathcal{R}_0$ in Figure~\ref{R0_L_alpha1}, where $\rho =0.8$ and all other parameters are given in Table~\ref{Tab1}. One sees that $\mathcal{R}_0$ increases quickly as $\alpha$ increases from $0$, hits the peak (which is higher, the lower is $L$), and then decreases slowly to the value obtained as $\alpha=1$ (no difference in attractiveness between infected individuals and the others).

Figure~\ref{R0_L_alpha1} also quantifies how the presence of odor-baited traps reduces part of the burden of mosquito bites for susceptible individuals so as to control the epidemics, showing that necessarily the effect will get smaller as the number of traps becomes larger.
As for the effect of $\alpha$, it can be seen that the value at which it reaches its peak increases as $L$ increases. If $\alpha$ has evolved under selective pressure on the virus to increase the attractiveness of infected individuals, a continuous presence of traps (or of alternative hosts) may change the selective pressures.

\begin{figure}[H]
    \centering
    \begin{tabular}{cc}
        a) & b)  \\
       \includegraphics[width=0.4\linewidth]{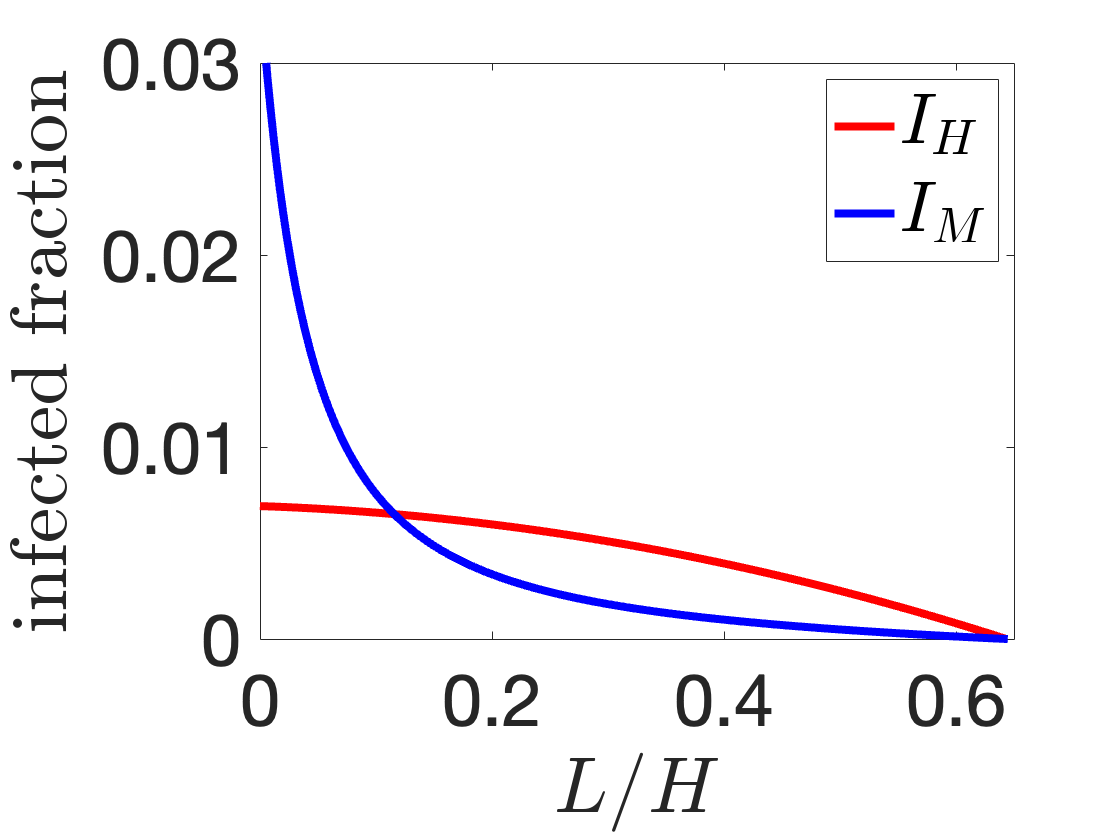}  & \includegraphics[width=0.4\linewidth]{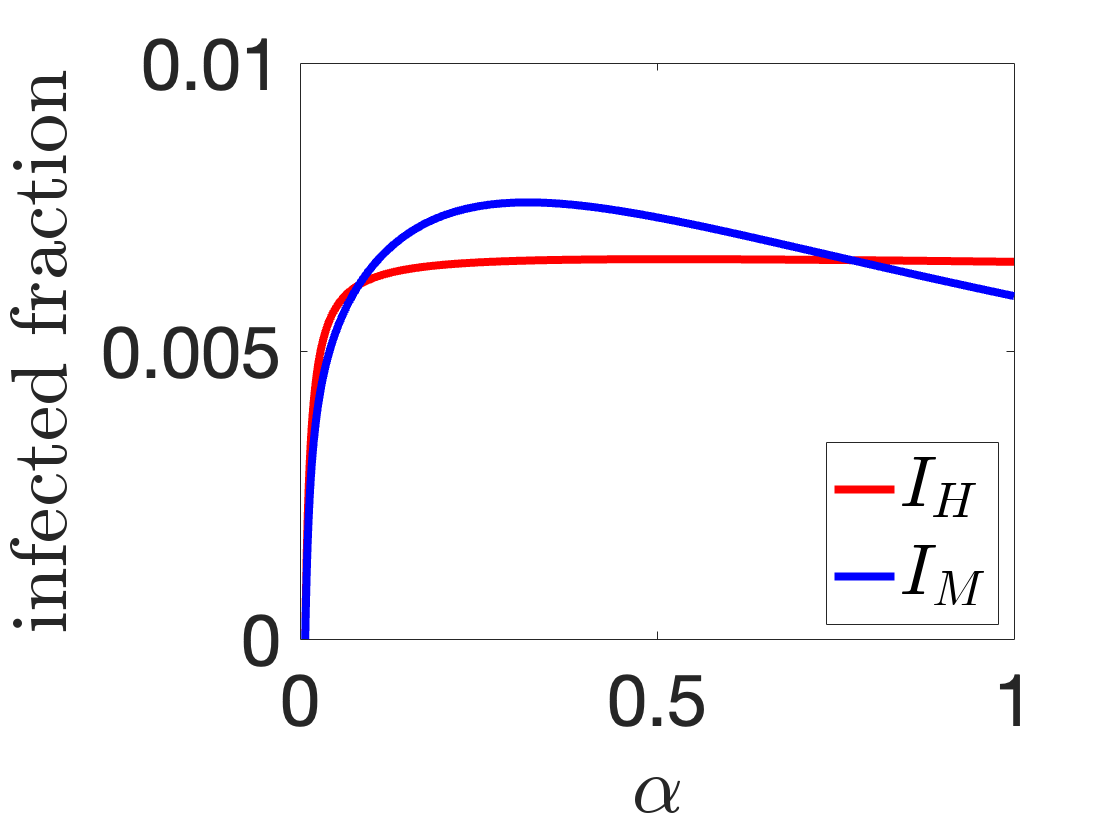}\\[0.5em] c) & d)  \\
            \includegraphics[width=0.4\linewidth]{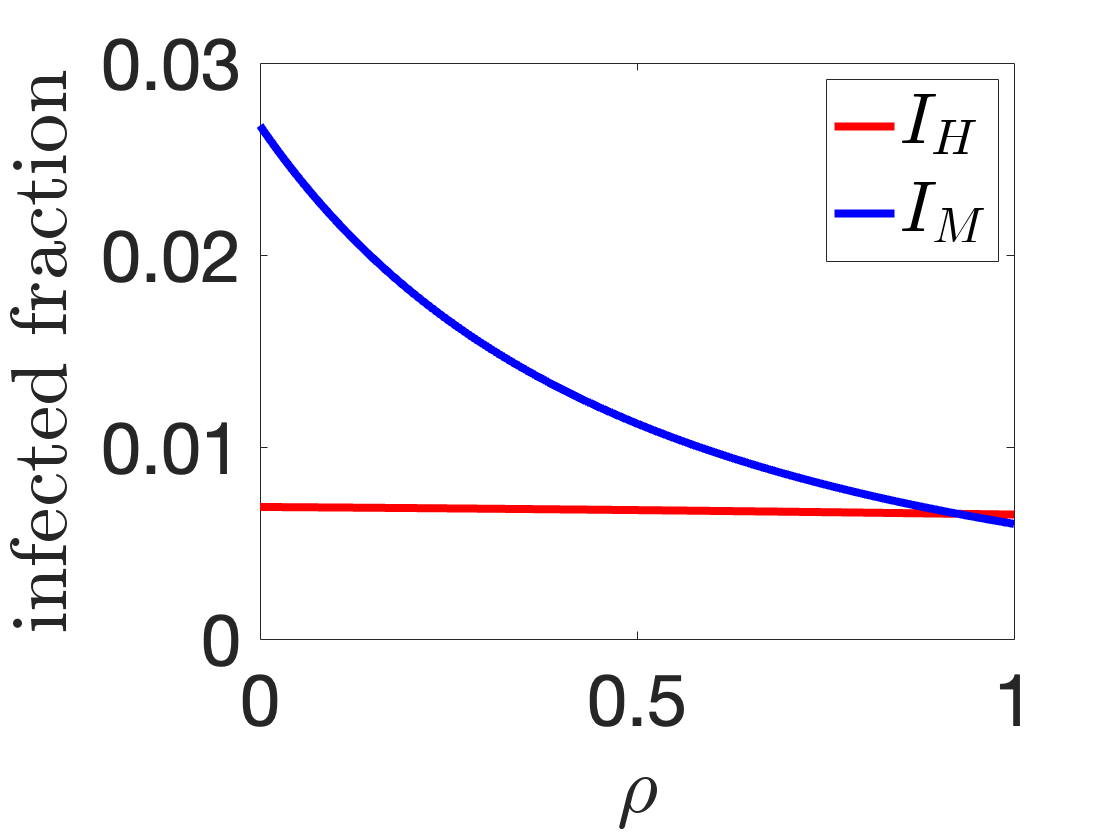}&
    \includegraphics[width=0.4\linewidth]{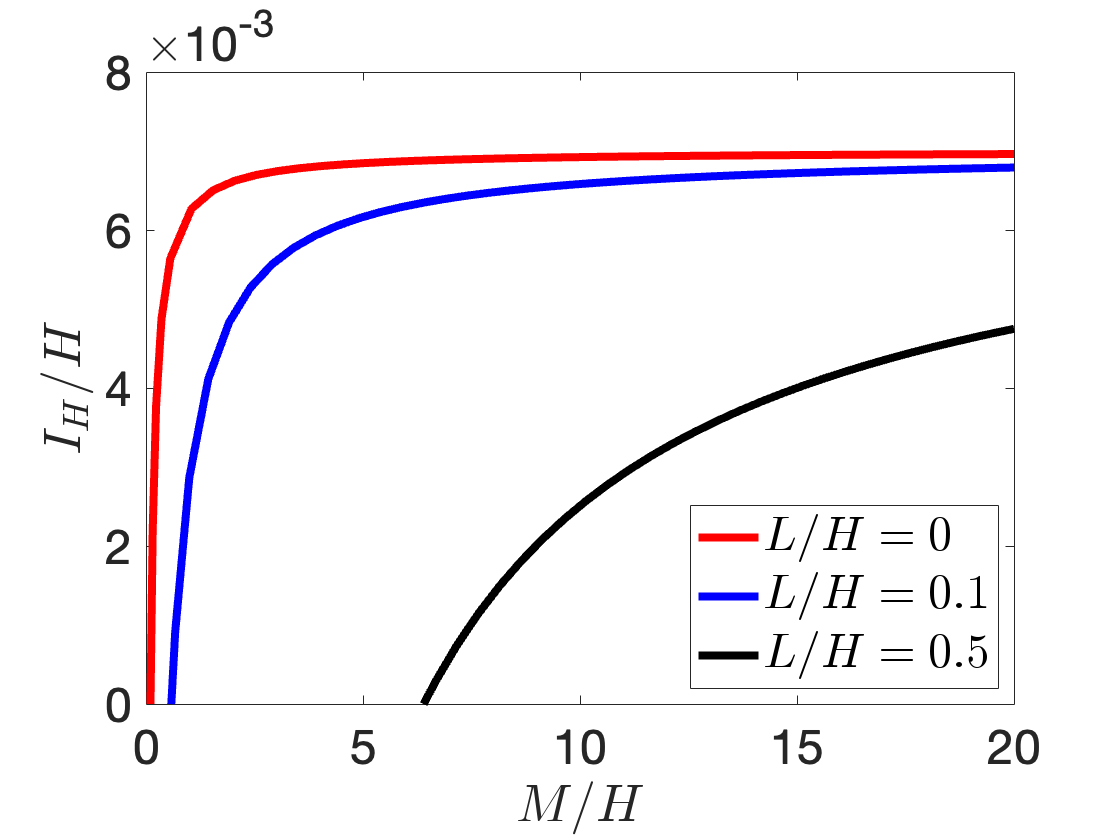}
\end{tabular}
     \captionsetup{width=14cm}
    \captionsetup{font=small}
   \caption{Panels a)--c): values of $I_M/\bar M$ and $I_H/\bar H$ at the endemic equilibrium as functions of $L/\bar H$, $\alpha$, or $\rho$, respectively. Panel d): value of $I_H/\bar H$ at the endemic equilibrium as a function of the vector to host ratio $\bar M/\bar H$ for three different values of $L$. Parameter values are listed in Table~\ref{Tab1}, while (unless otherwise stated) $L$, $\alpha$, and $\rho$ are kept at their reference values: $L = 1,000$, $\alpha =0.5$, $\rho = 0.8$. In panel a), $L$ varies between 0 and 6,436 when the epidemic goes extinct. In panel b), $\alpha$ varies between 0 (when the epidemic gets extinct) and 1. In panel c), $\rho$ varies between 0 and 1. }
    \label{fig_EE_vs_pars2}
\end{figure}

Besides $\mathcal{R}_0$, we also investigate the influence of the parameters on the endemic equilibrium.
In the first three panels in Figure~\ref{fig_EE_vs_pars2}, we show how the infected fractions of hosts ($I_H/H$) and of mosquitoes ($I_M/M$) vary with respect to $L$, $\alpha$, and $\rho$, while in the last panel, we show only the infected fractions of hosts ($I_H/H$) as a function of the vector-to-host ratio $M/H$.

The only case in which we do not see a monotone dependence on parameters can be seen in panel b), where both infected fractions reach a maximum at an intermediate value of $\alpha$, similarly to what is seen for $\Ro$. The other feature emerging from the Figure is that the dependence of
$I_H/H$ (the infected fraction of hosts at equilibrium) on the parameters is often extremely limited: this is true for $\rho$ (panel c) over its whole range, for $\alpha$ (panel b) and vector-to-host ratio (panel d), except that very close to the extinction point.

Finally, we wish to illustrate the occurrence of backward bifurcation as $\mathcal{R}_0$ varies around~$1$, according to the criterion presented in Theorem \ref{bifur}. As mentioned in Subsection \ref{sec32}, the quantity $M$ characterizing the backward bifurcation will always be negative if $\xi=0$, implying that in that case backward bifurcation cannot occur.
To illustrate the presence of a backward bifurcation in System~\eqref{endemic}, we then assume that the mosquito-borne disease has a high fatality rate $\xi = 0.04$. Other parameters are also adjusted to enhance the visibility of the backward bifurcation.

\begin{figure}[H]
    \centering 
    \captionsetup{width=14cm}
    \captionsetup{font=small}
    \includegraphics[width=0.8\linewidth]{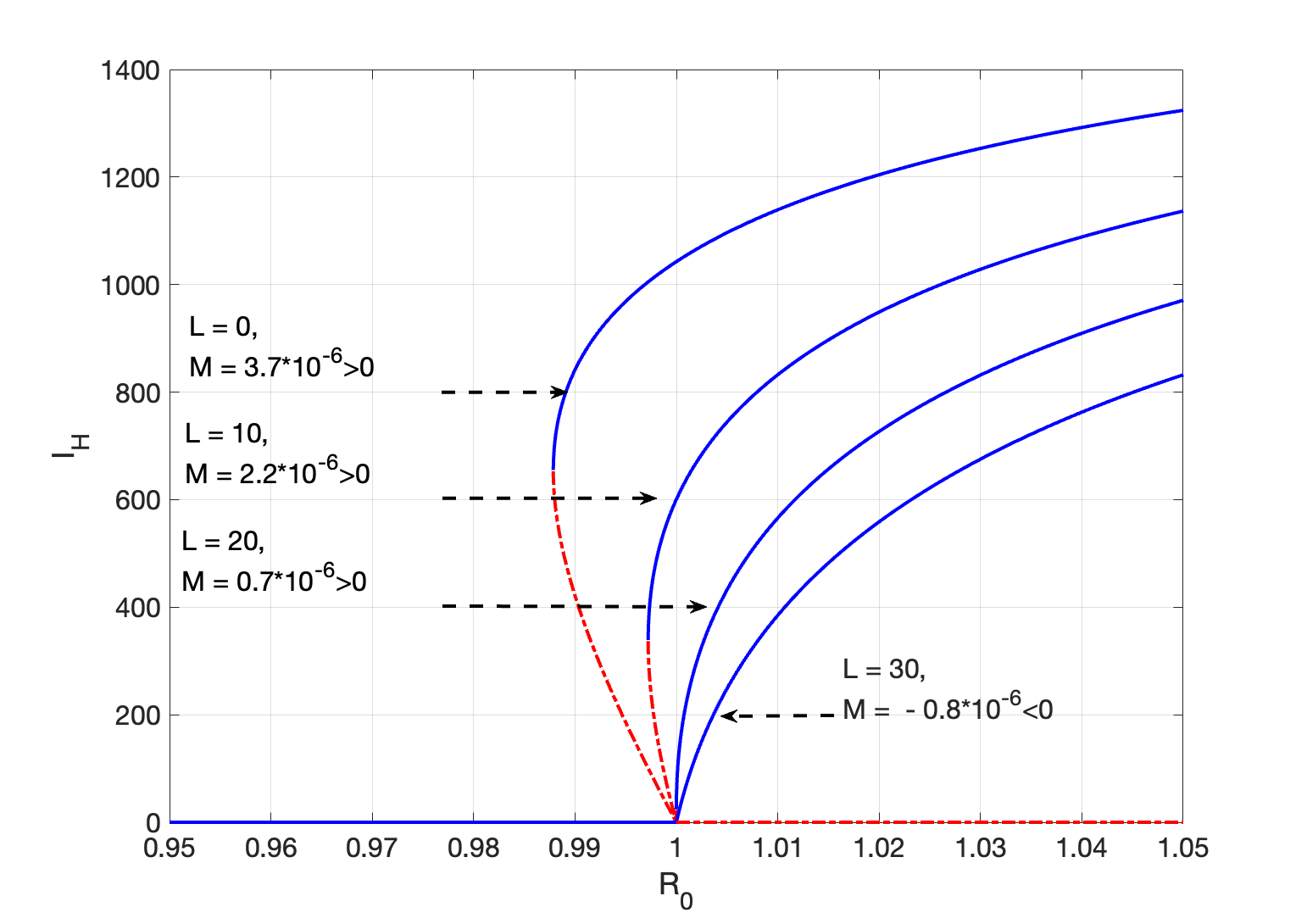}
    \caption{Bifurcation diagram of System~\eqref{endemic} with respect to  $\mathcal{R}_0$ around~$1$. Blue solid curves correspond to branches of asymptotically stable equilibria, red dashed ones to unstable ones. Parameters used are $\gamma = 0.01$, $\xi = 0.04$, $\rho = 0.4$, $\delta = 0.01$, $P_{hm}= 0.12$, $P_{mh}=0.22$, while other parameters are as in Tab.~\ref{Tab1}, except for $\alpha \in [0.1,0.15] $ to make $\mathcal{R}_0$ vary around~1.}
    \label{fig_bifur}
\end{figure}

In Figure~\ref{fig_bifur}, we then vary $\alpha$ between $0.1$ and $0.15$ to move $\mathcal{R}_0$ around 1 for $L = 0,\,10,\,20,\,30$. It can be seen that, if $L=0$ (no traps), the basic reproduction number is required to go below 0.99 for the endemic equilibrium to disappear, implying more effort of prevention is needed to control an epidemic. From another point of view, one expects to see a rather large infected fraction, even when the reproduction number $\mathcal{R}_0$ is just beyond the unit, a so-called epidemic burst.

The quantity $M$, appearing in the criterion for backward bifurcation, decreases as $L$ increases, becoming negative when $L$ goes beyond 20. Thus, the presence of odor-baited traps makes backward bifurcation gradually disappear.

\subsection{Simulation of an outbreak with deployment of traps}
\begin{figure}[H]
    \centering 
    \captionsetup{width=14cm}
    \captionsetup{font=small}
   \begin{tabular}{c|c}
      a)  & b) \\
      \begin{overpic}[width=0.4\linewidth]{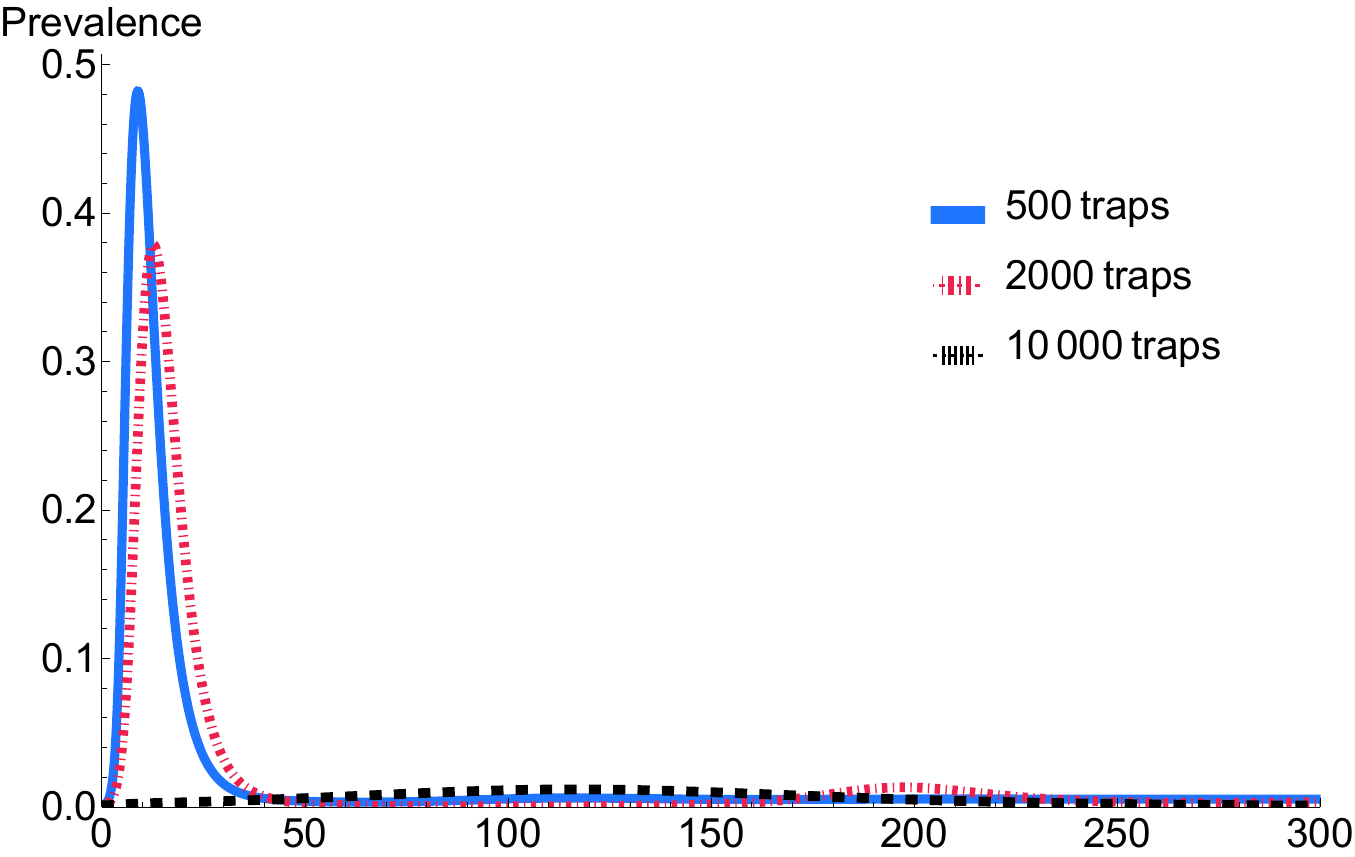}
      \put(90,-3){{\footnotesize $t$}}
      \end{overpic}
     &   \begin{overpic}[width=0.4\linewidth]{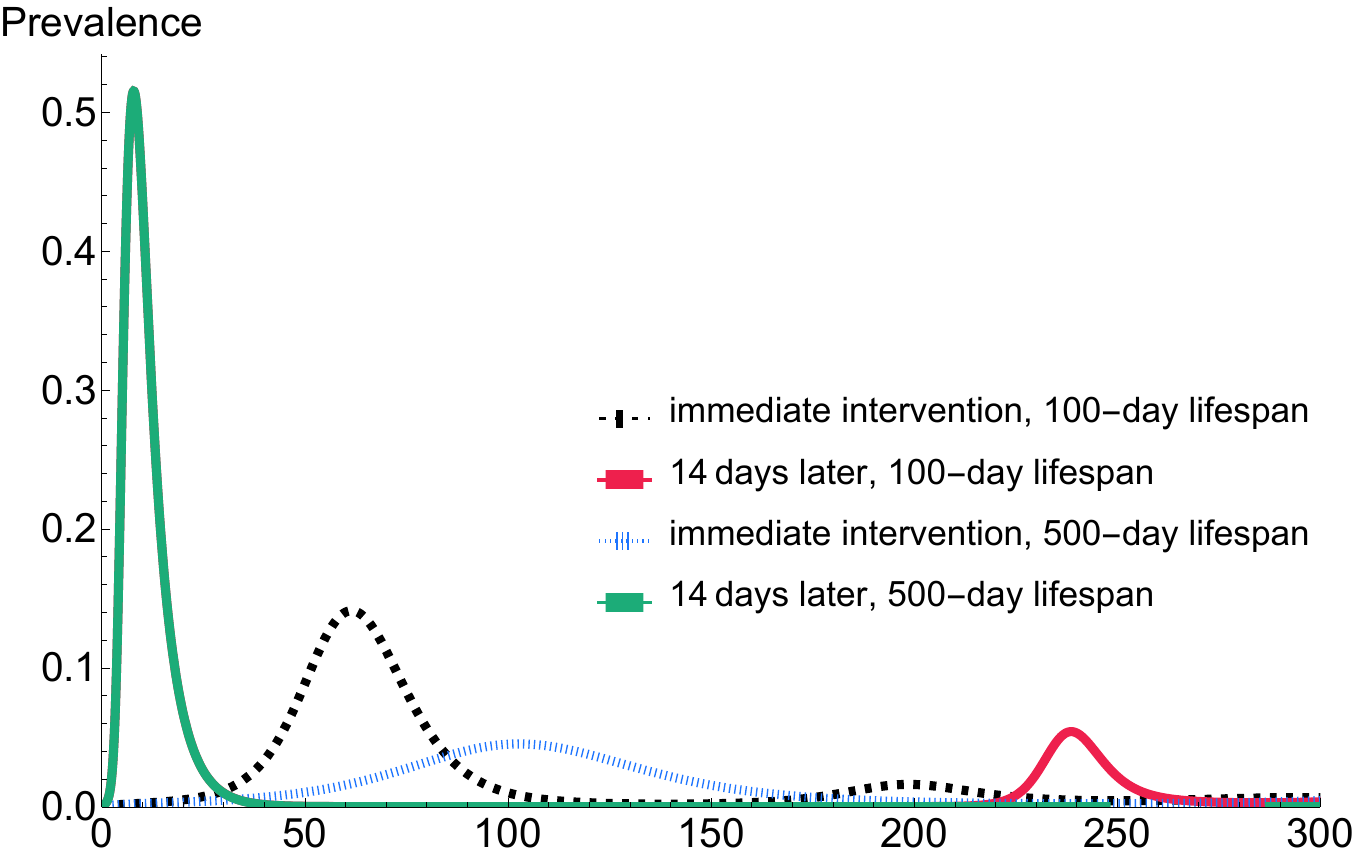}
     \put(90,-3){{\footnotesize $t$}}
      \end{overpic}
   \end{tabular}
    \caption{Prevalence plots showing proportion of $I_H$ over time. Parameters are from Table \ref{Tab1} with relative attractiveness~$\alpha = 0.5$, and trap lethality $\rho = 0.8. $  
    In the left panel, traps (with their effective number $L$ specified in the legend) are deployed at $t=0$ and work forever. In the right panel, traps (with $L=10,000$) are deployed either immediately (blue and black, dashed) or after 14 days (red and green, solid). For the black and red curves, it has been assumed that traps have an average lifespan of 100 days, while for the blue and green curves, an average lifespan of 500 days. Note that green and red curves overlap for all $t$ up to around 220.}
    
   \label{prev_plot}
\end{figure}

We now want to investigate the effect of odor-baited traps as a preventive/controlling measure to the outbreak. The results of numerical simulations are shown in Figure~\ref{prev_plot}.

In the left panel, we assume that traps have been deployed as a preventive measure, so they are present from the time ($t=0$) at which the outbreak starts. The Figure shows the outbreak dynamics for different values of deployed traps, implying that quite a large number of traps is essential for effective control. The small rebound that can be seen in the figure is due to the assumption that immunity wanes over time. 

In the right panel,  we assume that traps have been deployed as a control measure, and may be introduced either at the time ($t=0$) at which the outbreak starts, or $14$ days later.  We also assume that traps' efficacy declines exponentially over time, possibly because of a lack of proper maintenance. The figure shows how the delay in deploying the traps and the decay rate of their efficiency (expressed through the traps' effective lifespan) affect the simulation outcome. 

\section{Conclusion and discussion}\label{sect6}
The odor of hosts plays an important role during the host-seeking process of mosquitoes, serving as a biological tracking signal for mosquitoes to find the odor source.
Many studies~\cite{Takken1991,Dormont2021} indicate that mosquitoes depend on the host's odor totally in the long-distance search, partly during medium- and even short-distance search. Thus, it has been suggested that compounds attractive to mosquitoes may serve as the basis for complementary vector control strategies, such as those involving lure-and-kill traps or the development of mass trapping.

\andrea{
In this manuscript, we analyse the potential effect of the use of odor-baited lure-and-kill traps on the dynamics of a mosquito-borne infection, considering especially dengue or similar infections due to flavivirus. In this respect, it is important to remark that infected hosts may be more attractive to mosquitoes~\cite{Zhang2022}.

Building on the structure of Ross--Macdonald model~\cite{Tolle2009,Ross1910,Macdonald1957,Abboubakar2016}, the model 
includes three host compartments -- susceptible, infected, and recovered hosts, and two vector compartments--susceptible and infected mosquitoes. In the model, traps attract mosquitoes searching for hosts with two effects: the adult female population is temporarily reduced through the killing effect, but especially the biting rate on humans is decreased.

We established the basic properties (existence, non-negativity, and boundedness of solutions) of the model, computed the basic reproduction number $\mathcal{R}_0$, and proved the asymptotic stability of the disease-free equilibrium when $\mathcal{R}_0<1$.
As for positive equilibria, we showed that there exists a unique one when $\mathcal{R}_0>1$.
We also found that the system may have two endemic equilibria even though its basic reproduction number $\mathcal{R}_0$ is less than 1, due to backward bifurcation of the disease-free equilibrium at $\mathcal{R}_0 = 1$.  We obtained the exact condition for backward bifurcation (it occurs when a quantity $M$ is positive) and showed that $M$ may be positive only if the disease causes significant mortality in infected humans. Figure~\ref{fig_bifur} shows an example of backward bifurcation with two endemic equilibria existing for a range of $\Ro < 1$; it can be also seen that increasing the number of traps not only reduces the endemic level, but also makes the bifurcation forward.

When the disease causes no mortality and the odor-baited traps are harmless, the system becomes simpler, and we proved that the endemic equilibrium is locally asymptotically stable when $\mathcal{R}_0>1$.

The study was completed by a numerical study, using parameter values reasonable for dengue, and exploring especially the effect of the number of traps, of the mortality caused by traps, and of the differential attractiveness of infected hosts. As can be seen from the formula for $\Ro$, increasing the effective number of traps can reduce the prevalence at the endemic equilibrium down to zero. In Figure \ref{fig_EE_vs_pars2}, it is shown that more than 6,000 traps (counted in attractiveness equivalent of infected hosts) are needed for the eradication of an infection in a population of 10,000 hosts.

The effect of the relative (with respect to infected) attractiveness of not infected (susceptibles or recovered) on prevalence is not monotone. In the limit when susceptibles are not attractive at all, the infection would get extinct, as it would never be transmitted to new hosts; the maximum value of $\Ro$ depends on other parameter values and on the number of deployed traps, but is generally reached when their attractiveness is below 10\%, as transmission from infected hosts would get very efficient. It also decreases as their attractiveness becomes more similar to that of an infected individual. Certainly, there is an evolutionary pressure on the virus to increase the attractiveness of infected hosts, though not to extreme levels.

Finally, the level of the mortality of insects attracted to traps has a negligible effect on the long-term level of prevalence in hosts (see Figure \ref{fig_EE_vs_pars2}c).

We have also performed simulations of a control attempt of an outbreak through the deployment of traps. We have also assumed that, due to insufficient maintenance, the traps gradually lose attractiveness over time. We show that the outbreak control is possible if a sufficient number of traps is deployed as early as possible. The loss of attractiveness or immunity decay may lead to an epidemic rebound, but only on a rather long time scale.

Thus, we have seen that odor-baited traps may be relevant in controlling an outbreak of a mosquito-borne infection
if they are deployed in large numbers and if their attractiveness can be maintained long enough. In order for these results to be practical, one needs to quantify these conditions.
In the model, the number of traps needed is measured in terms of infected-equivalent. To be able to provide advice to public health, it would be necessary to estimate the relative attractiveness of traps and (possibly infected) humans. As far as we know, there are no studies in this direction.

Concerning the duration, it would probably be necessary to frequently restore the levels of attracting compounds in traps; to understand how frequently, the decay in attractiveness of traps should also be estimated.

Note that although we did not perform a formal sensitivity analysis, we also ran some simulations (not shown) with rather different parameter values. The outcome is qualitatively the same, but, as expected, quantitative results are rather different. 

Some of the assumptions are very restrictive. A very strong assumption is that, following the structure of Ross--Macdonald models, the mosquito life-cycle is neglected; it is assumed that there is a constant input of adult mosquitoes. It is instead likely that the presence of traps may reduce egg deposition, both because of the direct mortality and the reduced biting rate. In turn, a lower egg deposition may lead to a lower emergence of adults after the larval phase and thus to a smaller adult population; however, the number of adults that emerge may be influenced more by competition at the larval stage than by the level of egg deposition \cite{Marini2017a}.

We have also neglected the incubation periods and implicitly assumed that the length of the infectious stage in humans follows an exponential distribution. These unrealistic assumptions have been made to assess the role of traps and the differential attractiveness of infected hosts under the simplest assumptions. However, including these aspects does not change the structure of equilibria (except, possibly, for their stability), which is the main focus of this study. 

Moreover, the results have been obtained assuming that parameters are constant. However, mosquito demographic parameters, as well as biting rate and incubation period, are strongly affected by temperature and other environmental conditions~\cite{Delatte2009,Brugueras2020,Marini2020a}, which may vary even within a few weeks; quantitative predictions of the expected course of an outbreak should take these fluctuations into account. 

Another limitation of the study is the assumption that the population is homogeneous. In reality, individuals differ in exposure and attractiveness to mosquitoes~\cite{DeObaldia2022}; individuals will also differ in aspects of behaviour (such as use of repellents, air-conditioning, covering dresses) that affect their availability to mosquitoes~\cite{roosa2022}. Finally, heterogeneities in age and spatial location~\cite{Perkins2013} will affect the spread of an outbreak. We plan to investigate the impact of these aspects (including possible changes of behaviour in the presence of an epidemic outbreak~\cite{Buonomo2021}) in future work.
}
\appendix
\section{Proof of Theorem \ref{bifur}}
\label{app1}
As discussed in the main text, backward bifurcation at $\Ro=1$ occurs if and only if the quantity $M$, defined in~\eqref{M1M2}, is positive. 

Recall that $G$ and $H$ are defined, once a system has been written in the form~\eqref{eq_yz}.
In the case of System~\eqref{endemic}, $E_0 = [\bar N_{H}, 0,0,S_{M0},0]$,
\begin{equation}
    G = 
    \begin{pmatrix}
        -k_1& \beta_H \alpha \frac{S_{H}}{A_{t}}\\[0.5em]
        \beta_M  \frac{S_{M0}}{A_{0}}& -(\mu + \frac{b\rho L}{A_0})
    \end{pmatrix},
\end{equation}
\begin{equation}\nonumber
    H_z = 
    \begin{pmatrix}
        -d & \delta &0\\
        0& -\delta -d & 0\\
        0& 0& -\mu -\frac{b\rho L}{A_{0}}
    \end{pmatrix}, \quad 
    H_y = 
    \begin{pmatrix}
        0& \beta_H \alpha \frac{\bar N_{H}}{A_{0}}\\
        \gamma & 0\\
        -\beta_M  \frac{S_{M0}}{A_{0}}&0
    \end{pmatrix},
\end{equation}
with $A_{0} = L+ \alpha \bar N_{H}$ and 
\begin{equation}\nonumber
    w =\frac{1}{n} \left[1,\frac{A_{0}k_1}{\beta_M  S_{M0}}\right],\quad  v = \frac{1}{n} \left[1, \frac{A_{0} k_1}{\beta_H \alpha \bar N_{H}}\right]^{\rm T},\quad n = \sqrt{1+ \frac{A_{0}k_1}{A_{0}\mu + b\rho L}}.
\end{equation}
Therefore, we have 
\begin{equation}\nonumber
    U:=H_z^{-1}H_y v = \frac{1}{n}\left[\frac{\gamma}{d +\delta}+\frac{d+\xi}{d}, \frac{-\gamma}{d+\xi}, \frac{\beta_M S_{M0}}{A_{0}\mu + b \rho L}\right]^{\rm T},
\end{equation}
and we can compute $M_1$ and $M_2$.

Let us first calculate $M_1$ and in the light of~\eqref{M1M2}; defining
\begin{equation}\nonumber
    \begin{aligned}
        M_1(1,1) &:= w_1v_1 \sum_{k=1}^{2}\left(\frac{\partial  G_{11}(E_0)}{\partial y_k} + \frac{\partial  G_{1k}(E_0)}{\partial y_1}\right) v_k,\\
        M_1(1,2) &:= w_1v_2 \sum_{k=1}^{2}\left(\frac{\partial  G_{12}(E_0)}{\partial y_k} + \frac{\partial  G_{1k}(E_0)}{\partial y_2}\right) v_k,\\
        M_1(2,1) &:= w_2v_1 \sum_{k=1}^{2}\left(\frac{\partial  G_{21}(E_0)}{\partial y_k} + \frac{\partial  G_{2k}(E_0)}{\partial y_1}\right) v_k,\\
        M_1(2,2) &:= w_2v_2 \sum_{k=1}^{2}\left(\frac{\partial  G_{22}(E_0)}{\partial y_k} + \frac{\partial  G_{2k}(E_0)}{\partial y_2}\right) v_k,\\
    \end{aligned}
\end{equation}
we have  
\begin{equation}\label{M_1}
    M_1 = M_1(1,1)+M_1(1,2)+M_1(2,1)+M_1(2,2).
\end{equation}
Performing the computations, we obtain
\begin{equation}\label{M_1_cal}
    \begin{aligned}
        M_1(1,1) &= M_1(1,2) = -w_1v_1v_2\beta_H \alpha  \frac{\bar N_{H}}{A^2_{0}},\\
        M_1(2,1) &= -2 w_2 v_1^2 \beta_M ^2 \frac{S_{M0}}{A^2_{0}} + w_2 v_1 v_2  \frac{b \rho L}{A^2_{0}}, \\
        M_1(2,2) &= w_2 v_1 v_2  \frac{b \rho L}{A^2_{0}}.  
    \end{aligned}
\end{equation}

We next calculate $M_2$.
Similarly, through defining
\begin{equation}\nonumber
    \begin{aligned}
        M_2(1,1) &:= w_1v_1 \sum_{k=1}^{3}\frac{\partial G_{11}(E_0)}{\partial z_k}U_k,\; \quad
        M_2(1,2) := w_1v_2 \sum_{k=1}^{3}\frac{\partial G_{12}(E_0)}{\partial z_k}U_k,\\
        M_2(2,1) &:= w_2v_1 \sum_{k=1}^{3}\frac{\partial G_{21}(E_0)}{\partial z_k}U_k,\; \quad
        M_2(2,2) := w_2v_2 \sum_{k=1}^{3}\frac{\partial G_{22}(E_0)}{\partial z_k}U_k,\\
    \end{aligned}
\end{equation}
we have  
\begin{equation}\label{M_2}
    M_2 = M_2(1,1)+M_2(1,2)+M_2(2,1)+M_2(2,2).
\end{equation}
The computations yield
\begin{equation}\label{M_2_cal}
    \begin{aligned}
        M_2(1,1) &= 0,\\
        M_2(1,2) &= w_1v_2 \beta_H\alpha\left[\frac{U_1}{A_{0}} - \alpha\frac{(U_1+U_2)\bar N_{H}}{A^2_{0}}\right],\\
        M_2(2,1) &= w_2v_1\beta_M \left[\frac{U_3}{A_{0}} - \alpha\frac{(U_1+U_2)S_{M0}}{A^2_{0}}\right],\\
        M_2(2,2) &=  w_2v_2b\rho \alpha \frac{(U_1+U_2) L}{A^2_{0}}.
    \end{aligned}
\end{equation}

Combining~\eqref{M1M2} with~\eqref{M_1}--\eqref{M_2_cal}, we finally obtain 
\begin{equation}\label{M_fin}
    \begin{aligned}
        M &= M_1 - 2 M_2,\\
        M_1 &= \frac{2k_1}{n^3}\left[\frac{k_1 b \rho L }{\beta_H \alpha \bar N_{H} \beta_M  S_{M0}} - 2 \frac{1}{A_{0}} \right],\\
        M_2 &= \frac{k_1}{n^3}\left[\frac{1}{\bar N_{H}}\left(\frac{d + \xi}{d} + \frac{\gamma}{d+\delta}\right)- \frac{\alpha}{A_{0}}\frac{d+\xi}{d} + \frac{ \beta_M - \alpha \mu (d +\xi)/d}{\mu A_{0} + b \rho L} \right].
    \end{aligned}
\end{equation}
Using the condition (at $\tau=0$) $\Ro=1$, the first term in $M_1$ simplifies greatly, and one obtains
\begin{multline*}
    M =\frac{2k_1}{n^3}\bigg[\frac{b \rho L }{A_{0}(\mu A_{0} + b \rho L)}  +\left( \frac{\alpha}{A_{0}} + \frac{ \alpha \mu }{\mu A_{0} + b \rho L}\right)\left(1+\frac{\xi}{d} \right)\\
    -  \frac{2}{A_{0}} - \frac{ \beta_M }{\mu A_{0} + b \rho L} - \frac{1}{\bar N_{H}}\left(\frac{d + \xi}{d} + \frac{\gamma}{d+\delta}\right)\bigg]. 
\end{multline*} 
This expression can be rewritten as
\begin{equation}\label{M_xi_0}
    \frac{n^3}{2k_1}M = \frac{\xi}{d}\left( \frac{\alpha(2 \mu A_{0} + b \rho L)}{A_{0}(\mu A_{0} + b \rho L)} - \frac{1}{\bar N_{H}}\right) - C ,
\end{equation}
where 
 \begin{multline}\label{C}
    C =- \frac{b\rho L  + 2\alpha \mu A_{0}+\alpha b \rho L - 2  \mu A_{0}-2  b \rho L -  \beta_M A_{0} }{A_{0}(\mu A_{0} + b \rho L)}+ \frac{1}{\bar N_{H}}\left(1 + \frac{\gamma}{d+\delta}\right) \\
    =  \frac{( 1- \alpha)(2 \mu A_{0} + b \rho L)}{A_{0}(\mu A_{0} + b \rho L)} + \frac{ \beta_M}{\mu A_{0} + b \rho L} + \frac{1}{\bar N_{H}}\left(1 + \frac{\gamma}{d+\delta}\right) > 0,
\end{multline}
since $\alpha\le 1$.

Thus, we see that $\xi > 0$ is a necessary condition for backward bifurcation. To be more precise, $M > 0$ if and only if
\begin{equation*}
\frac{\xi}{d}\left( \frac{\alpha(2 \mu A_{0} + b \rho L)}{A_{0}(\mu A_{0} + b \rho L)} - \frac{1}{\bar N_{H}}\right) > C.
\end{equation*}
Notice that this condition is exactly the condition~\eqref{cond1_back} given in the statement of Theorem~\ref{bifur}.

A necessary condition for~\eqref{cond1_back} to hold is
$$
\frac{\alpha(2 \mu A_{0} + b \rho L)}{A_{0}(\mu A_{0} + b \rho L)} - \frac{1}{\bar N_{H}} > 0 \iff
\frac{\alpha \bar N_{H}(2 \mu \alpha \bar N_{H} +(2 \mu + b \rho) L)}{(\alpha \bar N_{H}+L)( \mu \alpha \bar N_{H} +( \mu + b \rho) L)} > 1,
$$
which is equivalent to
\begin{equation*}
    \frac{L}{\bar N_H} < \sqrt{\frac{\mu}{\mu+b\rho}}\alpha,
\end{equation*}
that is condition~\eqref{cond_back} of Theorem \ref{bifur}.

From the results in~\cite{Boldin2006}, the theorem follows.
\begin{center}
 \bf Acknowledegments   
\end{center}
This work was supported by MUR PNRR BaC PER INF-ACT S4 BEHAVE-MOD project (No. PE00000007, CUP: I83C22001810007). HZ was supported by the National Natural Science Foundation of China (Grant No. 12171396). 
The research visit of HZ to the University of Trento was funded by the China Scholarship Council (CSC) (Grant No. 202306990082). AP is a member of the ``Gruppo Nazionale per l'Analisi Matematica e le sue Applicazioni" (GNAMPA), while MS and CS are members of the ``Gruppo Nazionale per la Fisica Matematica" (GNFM) of the ``Istituto Nazionale di Alta Matematica" (INdAM).
\begin{center}
 \bf Disclosure statement   
\end{center}
The authors report there are no competing interests to declare.


 \bibliographystyle{plain}
 \bibliography{Biblio_VB}

\end{document}